\documentclass[letterpaper,12pt]{article}

\usepackage{mathptmx}

\usepackage{graphics}

\usepackage[authoryear,comma,sectionbib]{natbib} 

    \usepackage{amsmath, amsthm, amsfonts, dsfont, amssymb}
    \usepackage[colorlinks=true, linkcolor=blue, urlcolor=blue, citecolor=blue]{hyperref}
    \usepackage{booktabs} 
    \usepackage{array}
    \usepackage[dvips]{graphicx}
    \usepackage{multirow}
    \usepackage{caption}
    \usepackage{setspace}
\DeclareGraphicsExtensions{.pdf}
\graphicspath{{figures/}}

\begin{document}

\title{\LARGE Quantifying playmaking ability in hockey}
\author{\large Brian Macdonald 
\qquad Christopher Weld
\qquad David C. Arney 
\\ \\
Department of Mathematical Sciences and Network Science Center\\
United States Military Academy \\  
    West Point, NY $10996$ \\
}
\date{{\footnotesize\today}}

\maketitle

\begin{abstract}
It is often said that a sign of a great player is that he makes the players around him better.  The player may or may not score much himself, but his teammates perform better when he plays.   One way a hockey player can improve his or her teammates' performance is to create goal scoring opportunities.  Unfortunately, in hockey goal scoring is relatively infrequent, and statistics like assists can be unreliable as a measure of a player's playmaking ability.  Assists also depend on playing time, power play usage, the strength of a player's linemates, and other factors.  In this paper we develop a metric for quantifying playmaking ability that addresses these issues.   Our playmaking metric has two benefits over assists for which we can provide statistical evidence:  it is more consistent than assists, and it is better than assists at predicting future assists.   Quantifying player contributions using this measure can assist decision-makers in identifying, acquiring, and integrating successful playmakers into their lineups.
\end{abstract}


\newpage
\tableofcontents
\listoftables
\listoffigures

\section{Introduction}

A player contributes to team productivity in many ways.  Besides obvious contributions such as goal scoring and assists, a player's presence, tendencies, and other more difficult to quantify dynamics also factor into the fluid direction a game progresses. While direct involvement in goal scoring remains a significant indicator of player performance, it does not encompass all of their contributions to the team.  

In this paper we quantify a player's marginal contribution, decompose this into competitive and altruistic contributions, and use these measures to develop a playmaking metric.  All of these measures can be used to assess an individual's contribution to his team. In particular, the playmaking metric quantifies a player's ability to improve the productivity of his teammates.

\subsection{Motivation}

The dynamics of a competitive sports team are governed by a complex underlying framework of player attributes and group dynamics.  Generally speaking, a collection of great players is assumed to yield great results, and organizations fight constraints of both the availability of such talent, and the inherent costs associated with acquiring them, to grow rosters with maximum potential.  Identifying productive players is therefore critical to team management when drafting and trading players, and targeting free agents.  

While prospective player value can be assessed using individual statistics of past performance, these measures can carry inherent biases, and arguably do not represent a player's total on-ice contributions.  There is much more to a player's performance than the commonly referenced goals, assists, and plus-minus statistics.  Flaws in these measures include that they do not account for shorthanded and power-play usage, nor the strength of their team and teammates.  Scouts can assess player performance well but are unable to process every player in every game with the efficiency a computer has.  Developing advanced statistics that are capable of better quantifying player contributions is an aspiration for many, and it is the goal that we have here regarding a player's ability to improve his teammates' performance.

\subsection{Problem approach}

To assess a player's playmaking ability we first define the marginal contribution by quantifying a player's total productivity for his team.  This contribution is decomposed into components of competitive and altruistic contributions -- akin to goal scoring ability and remaining or ``other'' contributions.  Using a player's altruistic contribution, along with assists, we ultimately develop the playmaking metric.  

The playmaking metric accounts for the strength of a player's linemates, and is based on $5$-on-$5$ statistics so they are independent of how much power play or shorthanded time a player receives.  It is uses both shots and goals and is less subject to random fluctuations than metrics based only on goals.   We demonstrate that our metric is more consistent than assists by showing the year-to-year correlation of our metric is higher than that of assists.  We also confirm our metric is better than assists at predicting future assists by showing that we get a lower mean-squared error between predicted assists and actual assists when using our metric instead of assists. 

	\subsection{Previous applications}

Several previous studies in cooperative game theory provide the inspiration for our playmaking metric.  They have a common theme of identifying competitive and altruistic contributions of game participants, and explore both how to reasonably quantify these aspects, and the utility in doing so.  We discuss them next to provide the context of existing cooperative game theory approaches from which our analysis is derived.

Publications by \cite{cooperation-arney-peterson}, \cite{coop-peterson}, and \cite{coop-space-arney-peterson} address cooperation in subset team games.  In these games, team players pursue a common goal.  Each player has some positive contribution toward the goal.  Contributions are broken down into competitive and altruistic (selfish or unselfish, greedy or not greedy) components.  Using this decomposition, cooperation within organizations and teams is assessed for players or subsets of players.  
    
Determining competitive and altruistic player contributions has several useful applications such as assessing past and future performance and measuring chemistry among subsets of players.  By categorizing players according to their relative competitive and altruistic components, we can assess individuals and groups and conjecture the kinds of team compositions that lead to good group dynamics.  

	\paragraph{Pursuit and Evasion Games}

In pursuit and evasion games, a team of pursuers targets a team of evaders.  The pursuers attempt to catch the evaders before they reach a safe zone.  Each player operates autonomously.  

Using a na\"{\i}ve greedy search heuristic, a pursuer would chase its closest or most vulnerable target.  This approach represents completely competitive minded participants.  Alternatively, the pursuer could attempt to communicate with his or her teammates the location of the evader, and develop a more holistic strategy towards team success.  The pursuer may not get ``credit'' for catching that particular evader but did contribute to the success of the team.  In the latter circumstance, complementing competitive players with altruistic ones would intuitively lead to better results for the team.
    
\paragraph{Communications Networks}

\begin{figure}[h!] 
        \centering
             \includegraphics[width=.3\textwidth]{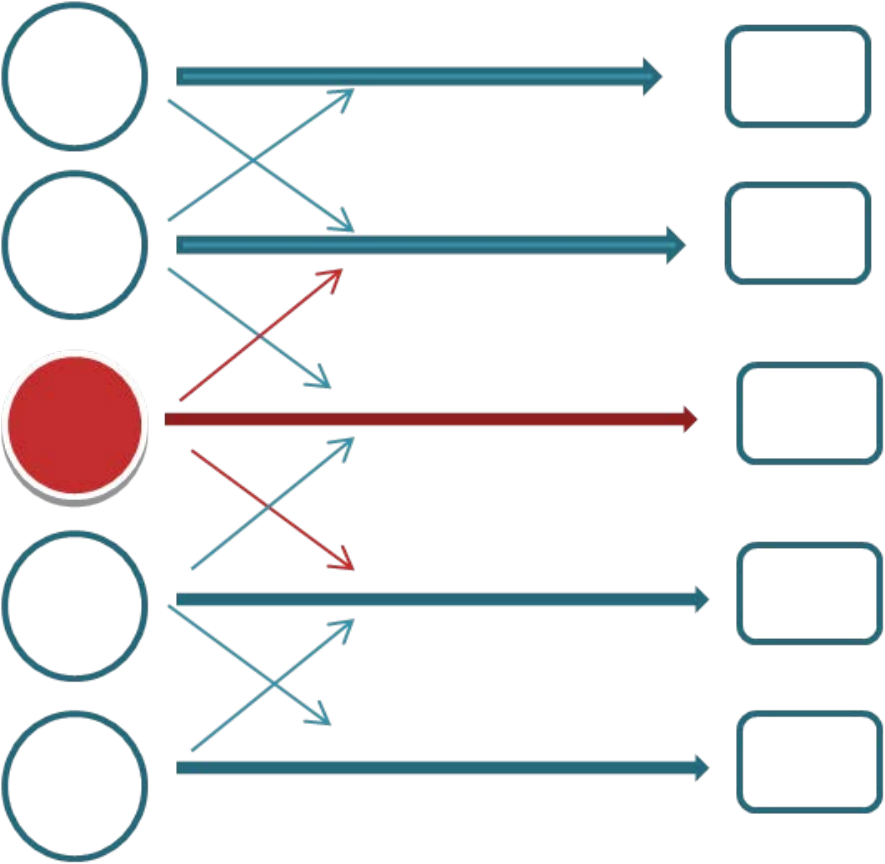}
             \caption{An illustration of a communications network}
		  \label{network-flow-example}
        \end{figure}

Consider an information network as in Figure \ref{network-flow-example}.
In this game, the players (nodes) on the left must transmit information through the network to the nodes on the right.  Players have different amounts of information to transmit, and each channel has a unique capacity.  The goal is to maximize the amount of information transmitted.  Nodes can transmit straight across their channel, or use the channels of adjacent nodes.  The nodes act autonomously, with only information about the nodes and channels next to them.  An optimum solution for the system is therefore unknown by the individual nodes.
    
Each player contributes to the goal through behaviors that could be classified as selfish or unselfish.  For example, a selfish player will transmit as much as possible in its own channel, while an unselfish player will let neighboring nodes with more information to transmit use their channel.  Different behaviors lead to different competitive and altruistic contributions for each player.  One can determine the combinations of player types, in terms of competitive and altruistic contribution, that lead to the best results.

\section{Notation and Definitions}\label{defs}
We will use notation consistent with definitions and results by Arney and Peterson in \cite{cooperation-arney-peterson}, \cite{coop-peterson}, and \cite{coop-space-arney-peterson} regarding cooperation in subset team games, beginning with the sets: 
\begin{eqnarray*}
&T:& \mbox{a set of all players on a given team,}\\
&A:& \mbox{a specific player or subset of players in $T$; in other words, $A \subseteq T$.}\\
&A^c:& \mbox{the complement of $A$; in other words, $A$'s teammates,}\\
	& & \mbox{or $T \backslash A$, the players in $T$ which are not in $A$.}
\end{eqnarray*}

The function $u$ is a utility function, or value function, which assigns a real number to every outcome of the game.  The quantity $u_X(Y)$ represents the value to $X$ when $Y$ participates.  The quantities we are most interested in, corresponding to the subsets $T$, $A$, and $A^c$, are
\begin{eqnarray*}\label{utt}
      &u_{T}(T):& \mbox{the value to the team, when everyone participates,}\\
      &u_{A^c}(T):&  \mbox{the value to everyone but $A$, when everyone participates, and}\\
      &u_{A^c}(A^c):&  \mbox{the value to everyone but $A$, when $A$ does not participate.}
\end{eqnarray*}
These definitions are revisited in greater detail when calculations are later completed.

\subsection{Defining and decomposing marginal contribution}

Our analysis begins with decomposing marginal contribution into its competitive and altruistic components, denoted as
\begin{eqnarray*}
      &c(A):& \mbox{the competitive contribution of $A$, and}\\
      &a(A):&  \mbox{the altruistic contribution of $A$.}
\end{eqnarray*}
Marginal contribution is defined as the sum of these respective contributions,
\begin{align} \label{marg}
m(A) = c(A) + a(A).
\end{align}

The competitive component is determined from direct contributions by $A$.  The term \textit{direct} refers to tallies (i.e. goals in hockey) towards team productivity attributed to $A$.  This competitive component is the difference in the value to $T$ and the value to $A^c$, when everyone participates:
        \begin{align}\label{defcomp}
            c(A) = u_{T}(T) - u_{A^c}(T).
        \end{align}
It may be helpful to think of the phrase ``value to'' as ``productivity of'', so that competitive contribution can be thought of as the difference in the productivity of $T$ and the productivity of $A^c$.  

The altruistic contribution of $A$ is the difference in the value to $A$'s teammates when $A$ does and does not participate, and is defined as 
        \begin{align}\label{defalt}
             a(A) = u_{A^c}(T) - u_{A^c}(A^c).
        \end{align}
In other words, $a(A)$ is the difference in the productivity of $A$'s teammates when $A$ does and does not play.  This measure is high when the contributions of $A$ are valuable to $A$'s teammates, or, in other words, when $A$ increases the productivity of $A$'s teammates.

Substituting equations \eqref{defcomp} and \eqref{defalt} into \eqref{marg}, we arrive at an equivalent expression for marginal contribution of $A$:
    \begin{align}
        m(A) &= c(A) + a(A) = \left[ u_{T}(T) - u_{A^c}(T) \right] + \left[ u_{A^c}(T) - u_{A^c}(A^c) \right] \nonumber
	\end{align}
or
    \begin{align}\label{defmarg}
        m(A) &=u_T(T) - u_{A^c}(A^c).      
    \end{align}
This expression says that marginal contribution of $A$ is the difference in the productivity of the team when everyone plays and the productivity of $A$'s teammates when $A$ does not play.

\section{Assessing a Player's Contributions in Hockey}\label{goals}

Having established necessary background, terms, and definitions, we extend this methodology to the sport of hockey.  Specifically, we identify a hockey player's marginal contribution and decompose that contribution into competitive and altruistic components.  Using these measures, we can subsequently use a players altruistic component to develop a measure of a player's playmaking abilities. 

	\subsection{Data}

Our analysis pulled data from \cite{nhlcom}.  Working with any professional sports data set provides inherent advantages and disadvantages, and hockey is no different.  Before proceeding with our analysis, it is worthwhile to highlight some of the pros and cons of working with this database.  

Conversely to a model parameterized entirely by its designer, analyzing competitive sports presents challenges unique to working with prescribed model parameters and subsequently produced data.  \cite{cooperation-arney-peterson}, \cite{coop-peterson}, and \cite{coop-space-arney-peterson} focused on developing theories about cooperation, and applying it to pursuit and evasion games and information networks.  As theoretical scenarios, the designer has full control over all parameters.  Player attributes are adjusted to create circumstances of interest to the model's designer.  Similarly, they also control the rules of the game, and alter them accordingly to set their desired conditions.  
    
With exception of those connected to team management, a hockey analyst has no control over the kinds of players that play together and cannot try combinations of their choosing.  There is, however, an abundance of real data available for analysis.  Our focus is to identify useful data and choose appropriate, meaningful, and interpretable values and payoff functions.
    
Hockey data is conducive to analysis for several reasons.  The data is relatively accurate, complete, and detailed.  The NHL data used provided the players on the ice for every second of every game, and all corresponding events such as goals, shots, hits, giveaways, etc.  Sports data also provides easily quantifiable natural objective outcome values, such as goals scored or wins, that are not a subjective assessment.  In most other kinds of organizations, the ideas of value, outcomes, contributions, and teamwork are typically more subjective in terms of both measurement and definition.
                
\subsection{Defining contributions using goals} \label{gsa}
The next few sections walk through improving iterations of our analysis of player marginal contribution, and its decomposition into competitive and altruistic components.  Detailing this evolution is useful to illustrate the pitfalls of other seemingly simpler or more intuitive approaches, and reinforce the validity of our final solution.

We start quantifying contributions in hockey with perhaps the simplest choice for value, goals scored.  The value of a season to a subset of players is defined as goals scored by those players during that season. The quantity $u_X(Y)$ is defined as the goals scored by players in $X$ when the players in $Y$ participate.  We consider the case when $A$ is a single player, and we have 
        \begin{align*}
            u_{T}(T) &= \mbox{goals scored by the team when everyone participates,} \nonumber \\
            u_{A^c}(T) &= \mbox{goals scored by $A$'s teammates when everyone participates, and} \nonumber \\
            u_{A^c}(A^c) &= \mbox{goals scored by $A$'s teammates when $A$ does not play.}
        \end{align*}
Note that when we say ``when everyone participates,'' we do not mean the whole team is participating at the same time.  In hockey this never happens, since at most five players (plus a goalie) play at once for a given team.  Therefore, we interpret the scenario for everyone participating as any subset of five players in $T$ on the ice at a given time. Similarly, when we say ``when $A$ does not play'' we mean when a subset of $A^c$ is on the ice.

Here, and throughout this paper, we consider only 5-on-5 situations in which both goalies are on the ice.  We do not want our metrics to depend on if a player's coach happens to give him power play or short handed time, or happens to play him at the end of the game when one team has pulled their goalie.

    Let $G$ be the goals scored by $T$, let $g$ be the goals scored by player $A$, and let $gf$ (goals for) be the goals scored by the team when $A$ is on the ice.  Then we have 
         \begin{align*}
             u_T(T)         &= G \\
             u_{A^c}(T)     &= G - g \\ 
             u_{A^c}(A^c)   &= G - gf 
         \end{align*}
    Recalling that $c(A) = u_T(T)  -  u_{A^c}(T)$ and $a(A) =  u_{A^c}(T) - u_{A^c}(A^c)$, we have 
        \begin{align*} 
            c(A) &= G - (G - g) = g \\
            a(A) &= (G-g) - (G - gf) = gf - g
        \end{align*}
    With these definitions, player $A$'s competitive contribution, $c(A)$, is simply the goals scored by player $A$, and his altruistic contribution is the goals that his teammates score when he is on the ice.  Note that $a(A)$ is high when the team scores many goals when $A$ plays, but $A$ himself does not score many of the goals.  The team does well when $A$ plays, but $A$ is not necessarily getting the credit for the goals.

\begin{table}[h!]
\begin{center}
\caption{Top five forwards in altruistic contribution}
\label{cooptop5}
{\small
\begin{tabular}{llrrrrrrrrr}
  \addlinespace[.3em] \toprule 
Player & Pos & Team & $u_T(T)$ & $u_{A^c}(A^c)$ & $m(A)$ & $c(A)$ & $a(A)$ & A & Pts & Mins \\ 
  \midrule 
  H. Sedin   & C  & VAN & 162 & 95  & 67 & 9  & 58 & 44 & 53 & 1194 \\ 
  J. Toews   & C  & CHI & 167 & 98  & 69 & 20 & 49 & 27 & 47 & 1187 \\ 
  D. Sedin   & LW & VAN & 162 & 93  & 69 & 22 & 47 & 35 & 57 & 1145 \\ 
  R. Getzlaf & C  & ANA & 134 & 79  & 55 & 8  & 47 & 32 & 40 & 1114 \\ 
  B. Boyes   & RW & STL & 162 & 106 & 56 & 9  & 47 & 30 & 39 & 1146 \\ 
   \bottomrule 
\end{tabular}
}
\end{center}
\end{table}

The top five forwards in altruistic contribution, $a(A)$, are given in Table \ref{cooptop5}.  The last three columns denote assists (A), points (Pts), and minutes played (Mins) during the 2010-11 season.  The results are what we might expect. Players who play on good offensive teams and get a lot of assists, and may or may not score many goals themselves, have high altruistic contributions.  

\begin{figure}[h!]
            \begin{center}
             \includegraphics[width=.45\textwidth]{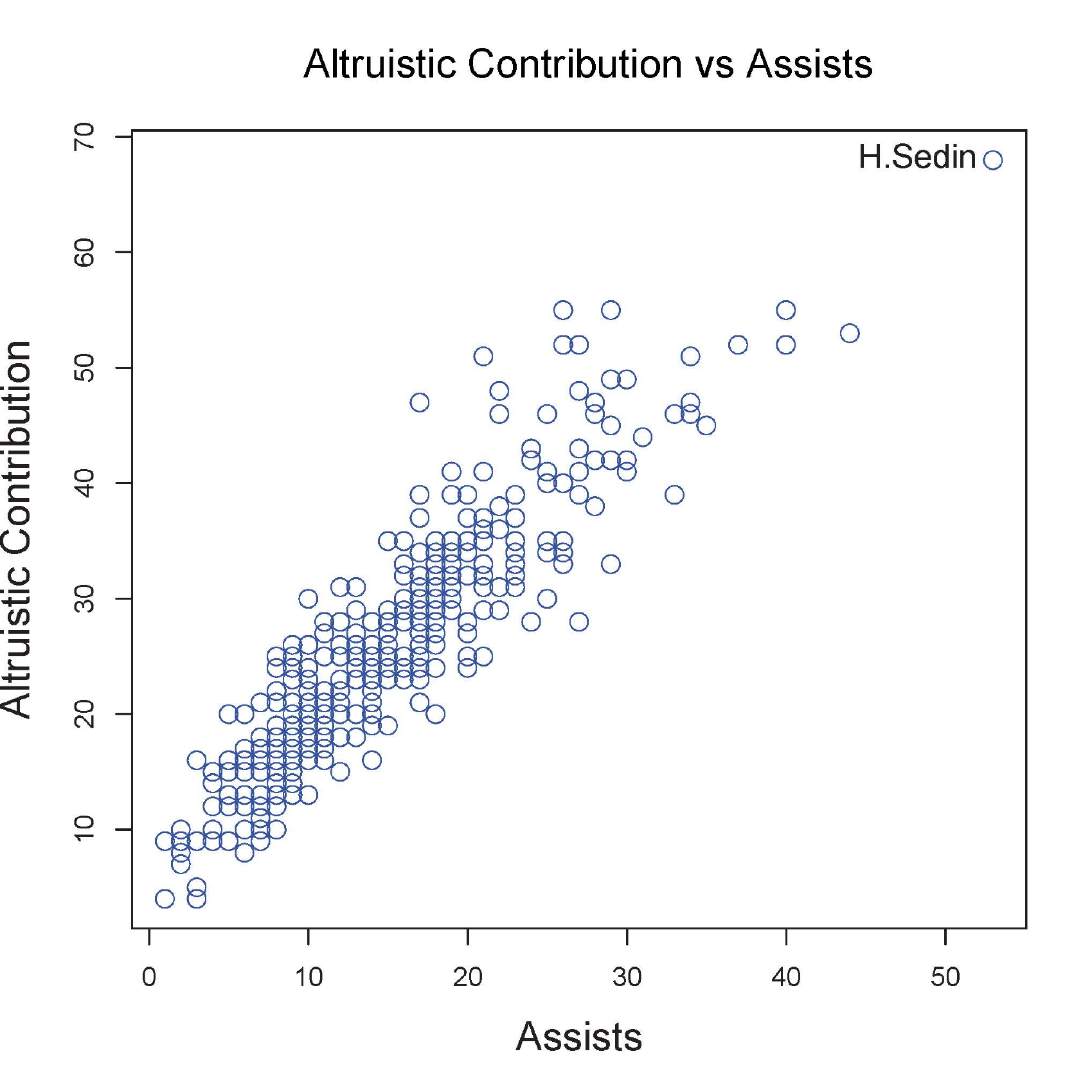}
            \includegraphics[width=.45\textwidth]{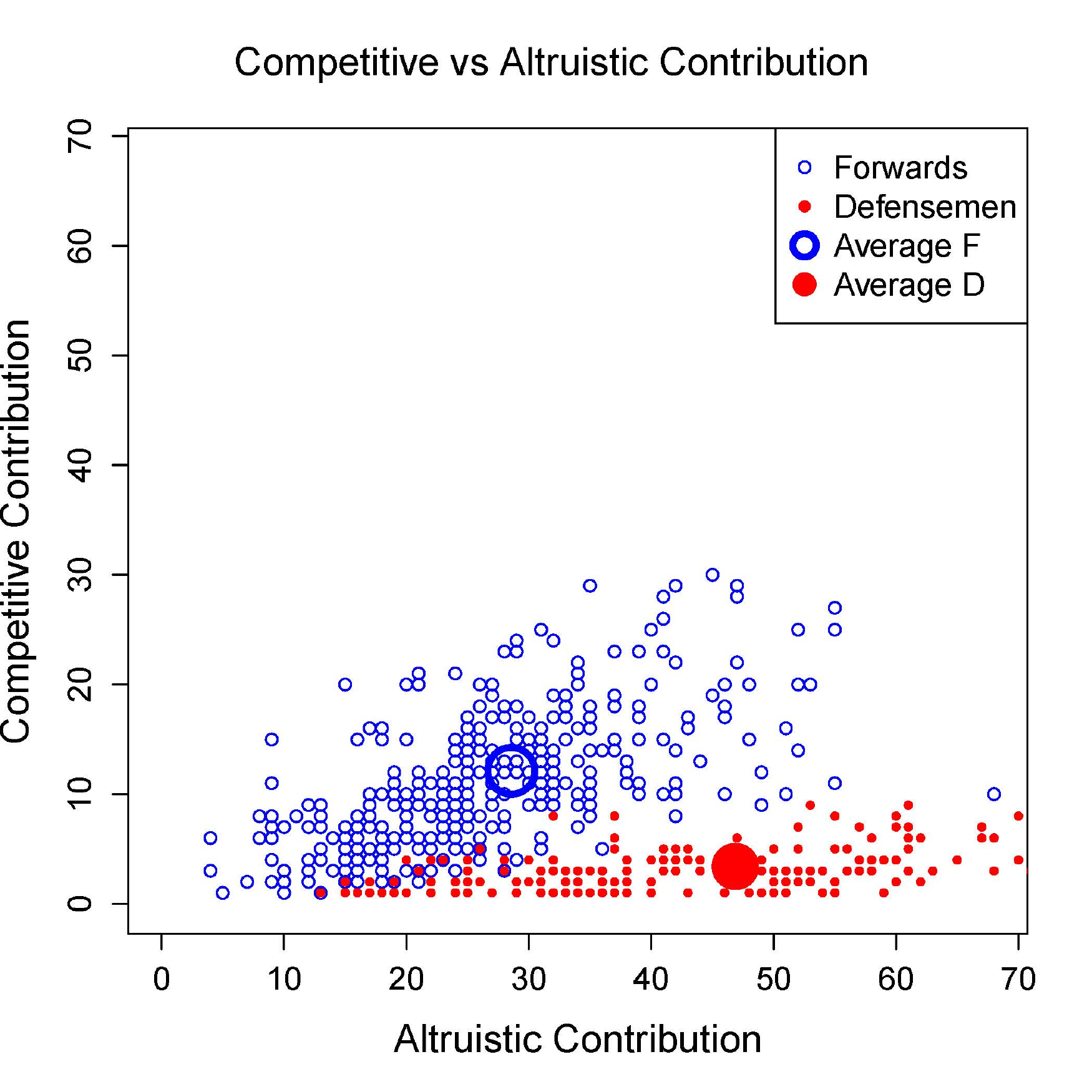}
            \caption{(Left) Scatter plot of assists vs altruistic contribution. (Right) Scatter plot of competitive versus altruistic contribution for forwards (blue circles) and defensemen (red dots).}
            \label{fig-goals}
            \end{center}
        \end{figure}

The correlation between assists and altruistic contribution is fairly high (0.91), and a scatterplot of altruistic contribution versus assists is given in the left of Figure \ref{fig-goals}.
In the right of Figure $\ref{fig-goals}$, we see how the competitive and altruistic contributions of both forwards and defensemen are distributed.   A distinct clustering of forwards and defensemen appears.  This is an intuitive result considering the inherent goal scoring opportunities (or lack thereof) that accompany their positions.  For example, defensemen typically have a low competitive contribution due to their relatively low goal scoring rate.  Defensemen also typically play more minutes than forwards, yielding higher marginal contributions, and therefore higher altruistic contributions.  

\subsection{Defining contributions using goals per 60 minutes}\label{gsa60}

Under the approach in Section \ref{gsa}, goals, assists, and altruistic contribution, for example, are highly influenced by playing time.  We expect a player receiving significant playing time to amass more opportunities for goals and assists, both of which influence competitive and altruistic measures.

To eliminate this playing time bias, we could alternatively capture statistics as a rate per 60 minutes.  This adjustment not only standardizes comparison between players, but mitigates the significant correlation between altruistic contribution and assists reflected in Figure \ref{fig-goals}.  Although assists and altruistic contribution remain correlated using a rate statistic, the correlation is not as extreme.

The simplest way to do this is to replace ``goals'' with ``goals per 60 minutes'' in the previous definitions.  Variables of interest are altered as follows:
        \begin{description}
            \item[$u_{T}(T)$] = goals per 60 minutes scored by the team during the times when $A$ does and does not play.
            \item[$u_{A^c}(T)$] = goals per 60 minutes scored by everyone except $A$ during the times when $A$ does and does not play.
            \item[$u_{A^c}(A^c)$] = goals per 60 minutes scored by everyone except $A$, during only the times when $A$ does not play.
        \end{description}
        \begin{eqnarray*}
                m(A) &=& \mbox{the difference in the goals per 60 minutes scored by the team} \\ 
					& &\mbox{during all times and during only the times when $A$ does not play.}\\
                c(A) &=& \mbox{the number of goals per 60 minutes player $A$ scored.}\\
                a(A) &=& \mbox{the difference in the goals per 60 minutes scored by $A$'s teammates } \\
					& &\mbox{during all times and during only the times when $A$ does not play.} 
         \end{eqnarray*}
We can still decompose $m(A)$ into two components:
    $$ m(A) = c(A) + a(A). $$
In other words, a player's marginal contributions are divided into two components, the goals per 60 minutes he scores, and the increase (or decrease) in the goals per 60 minutes his teammates score during all times and during only the times when $A$ is on the ice.

\subsection{A further adjustment to our definitions} \label{linemates}

A high altruistic contribution with the current definition indicates a player's team scored a lot of goals when they were on the ice relative to when they were off the ice, but they themselves did not score many of those goals.  Although it is tempting to associate a high altruistic rating with the innate playmaking qualities of the individual, this metric can prove misleading.  Consider the hypothetical team of 4 forward lines and 3 defense pairings in Table \ref{faketeam}.
    \begin{table}[h!]\centering
    \caption{A hypothetical team of above average and below average players.}
    \begin{tabular}{llllll}
        \addlinespace[.5em]
        \toprule
        LW  &   C   &   RW  & \quad  &   D   &  D  \\
        \midrule
        \textbf{Player A}  &  Above & Above  & \quad & Above  & Above  \\
        Below  &  Below & Below  & \quad  & Below  & Below  \\
        Below  &  Below & Below  & \quad  & Below  & Below  \\
        Below  &  Below & Below  & \quad  &  & \\
        \bottomrule
    \end{tabular}
    \label{faketeam}
    \end{table}

Player $A$, the below average player in the first line, typically plays with above average players on a team with mostly below average players.  He will have a high $u_{A^c}(T)$ because his teammates score a lot of goals when he plays (which he often did not contribute to because he is a below average player), and will have a low $u_{A^c}(A^c)$ because his team does not score that many goals when he does not play.  He will have a high $a(A)$ and categorize as ``unselfish."  Player $A$, however, may or may not have anything to do with the increase in goals when he plays, because the strength of his linemates is much greater than the strength of the rest of the team.  

Player $A$'s altruistic contribution should perhaps not be so heavily dependent on the performance of players he never plays with.  This observation motivates modified value and payoff functions that account for the strength of teammates a player experiences ice-time with.  The $u_{A^c}(A^c)$ term is calculated as a weighted average based on playing time with $A$, instead of as an unweighted average of team goals per 60 minutes scored when $A$ is off the ice.  

We note that this version of marginal contribution is similar to the With Or Without You (WOWY) and on-ice/off-ice statistics described in 
\cite{fyffe-vollman}, 
\cite{boersmawowy}, 
\cite{seppa}, 
\cite{gabewowy}, 
\cite{deltasot}, 
\cite{tango-wowy}, 
\cite{wilson-wowy}, and 
\cite{davidjohnson}, although some of those metrics use different data or are computed in slightly different ways.  Instead of writing 
    $$ m(A) = u_T(T) - u_{A^c}(A^c), $$
we could use the notation $GF_{on}$ for $u_T(T)$, and $GF_{off}$ for $u_{A^c}(A^c)$, and using \eqref{defmarg} we can write marginal contribution in a notation that is closer to what the online hockey analyst community would use:
    $$ m = GF_{on}- GF_{off}.$$
This notation is perhaps more intuitive and highlights that marginal contribution is measuring what happens when a player is on the ice versus off the ice.  We will continue to use $GF_{on}$ and $GF_{off}$ in lieu of $u_T(T)$ and $u_{A^c}(A^c)$ going forward, especially since we have changed the meaning of these terms slightly. 

The first term $GF_{on}$ is simply the goals per 60 minutes scored by the team when $A$ is on the ice.  For the second term, we first let $GF_i$ be the goals for per 60 minutes for player $i$ when playing \textit{without} $A$, and let $w_i$ denote playing time \textit{with} player $A$.  Then we define $GF_{off}$ to be the weighted average
    $$ GF_{off} = \frac{\sum GF_i \, \, w_i}{\sum w_i}, $$
where the sums are taken over all $i$.  Teammates frequently paired with $A$ have high $w_i$ and are more influential in this statistic, while those never playing with $A$ will have no affect on $GF_{off}$.  This prevents undue influence from teammates that $A$ is seldom or never paired with, and likewise emphasizes data with greater supporting information.

We remark that we could have chosen to define marginal contribution using one of the regression-based metrics referenced in the Section \ref{conclusions and future work}.  We have chosen the method presented here because of speed of computation, and because we would ultimately like to consider the case where $A$ is a subset of two or more players instead of a single player.  In Table $\ref{M5}$, we see that our choice for marginal contribution quantifies performance well, as these players are generally regarded as being among the best offensive players in the league.
        \begin{table}[h!]
        \begin{center}
        \caption{Top five forwards in marginal contribution using goals per 60 minutes} \label{M5} {\footnotesize \begin{tabular}{rrrrll}
          \addlinespace[.3em] \toprule
         Rk& Player & Pos & Team & m & Time \\
          \midrule
          1 & Sidney Crosby & C & PIT & 1.55 & 3614 \\
          2 & Henrik Sedin & C & VAN & 1.28 & 4530 \\
          3 & Pavel Datsyuk & C & DET & 1.27 & 4259 \\
          4 & Daniel Sedin & LW & VAN & 1.22 & 4164 \\
          5 & Alex Ovechkin & LW & WSH & 1.17 & 4896 \\ 
           \bottomrule
        \end{tabular}
        }
        \end{center}
        \end{table}

\begin{figure}[h!]         
                \begin{center}
                \includegraphics[width=.45\textwidth]{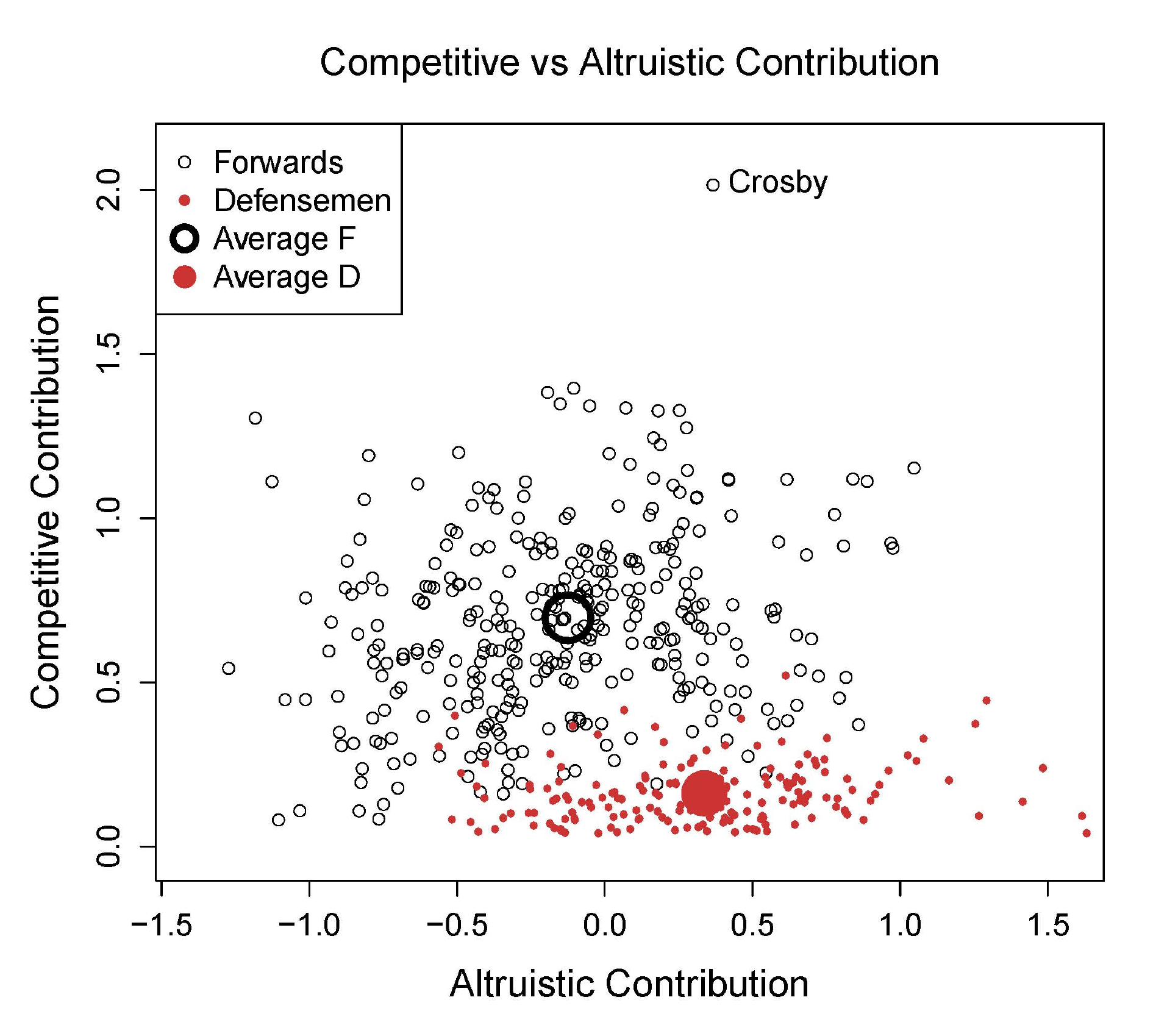}
                \includegraphics[width=.45\textwidth]{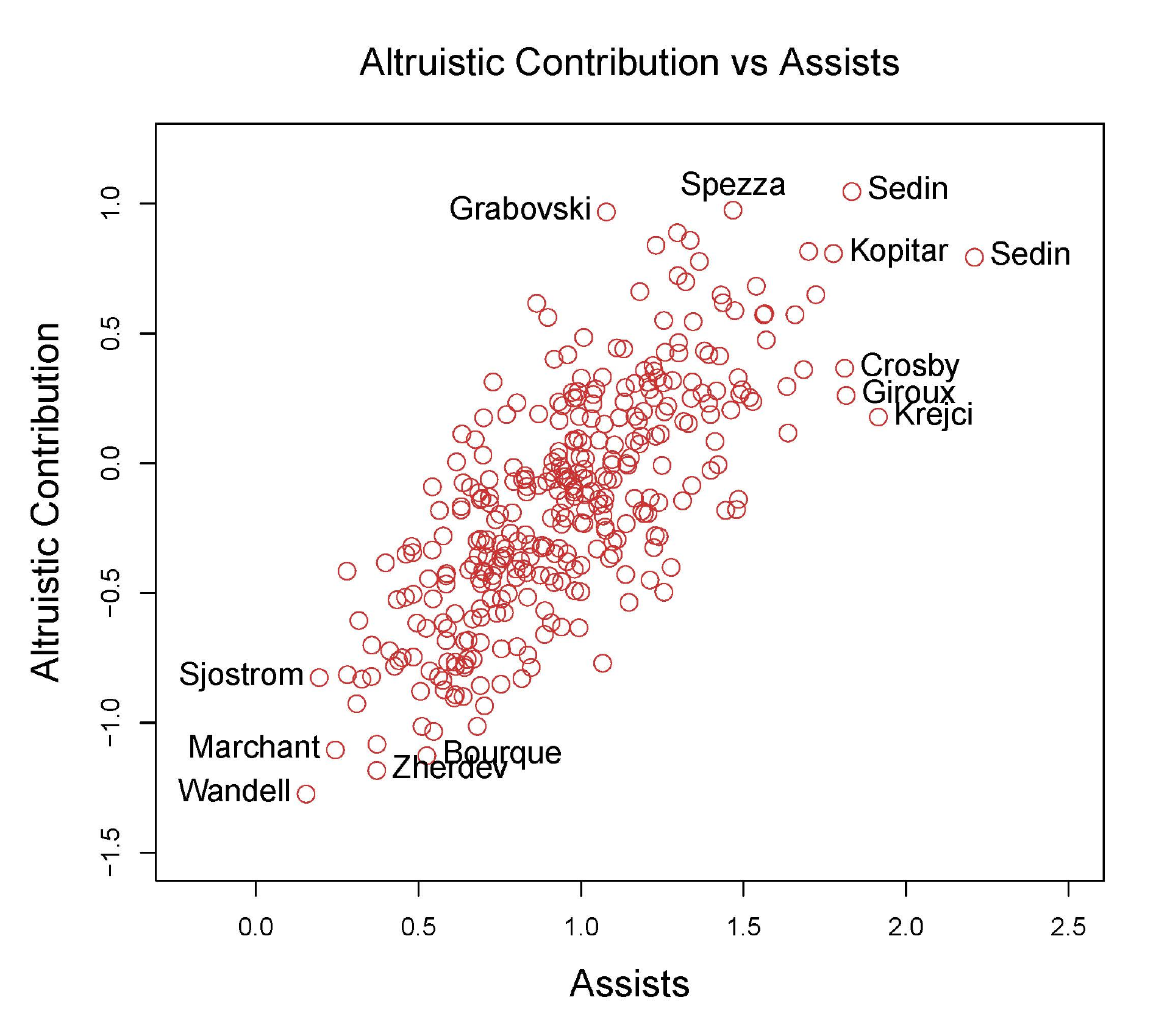}
                \caption{(Left) Competitive versus altruistic contribution for forwards (black circles) and defensemen (red dots) for 2010-11.  (Right) Altruistic contribution versus assists per 60 minutes for forwards in $2010$-$11$.}  \label{assists-alt-g60w}   
                \end{center}
            \end{figure}

This new definition varies a bit from those introduced in previous sections, but we can still decompose marginal contribution into competitive and altruistic components.  Player $A$'s competitive contribution $c(A)$ is the goals per 60 minutes scored by $A$ himself, and his altruistic contribution is everything else:
    $$ a(A) = m(A) - c(A) .$$
In the left of Figure \ref{assists-alt-g60w}, we see that competitive and altruistic contributions appear uncorrelated, especially for forwards (black circles), evidence that they are measuring different skills.  Sidney Crosby is an outlier, but this is not terribly surprising, especially since he only played half the season in $2010$-$11$ and we have a relatively small sample size in his case.  
In the right of Figure $\ref{assists-alt-g60w}$, we see that for forwards altruistic contribution is fairly correlated with assists per 60 minutes (correlation $\approx 0.75$).  
     
\subsection{Defining contributions using shots per 60 minutes}

Since scoring is relatively infrequent in hockey, goals can be somewhat unreliable to consistently represent on-ice performance. 
In \cite{possessioniseverything}, \cite{shots-fwick-corsi}, and \cite{spm}, the authors conclude that shots are both more consistent than goals and better than goals at predicting future goals.  Since assists are based on goals, a player's assists are subject to the same randomness.  Our previously developed altruistic contribution is based on goals and has the same problem.  In fact, it actually has a slightly lower year-to-year correlation than assists. 

Unfortunately, while the NHL records assists, or the number of a player's passes that immediately precede a teammate's goal, they do not record the number of a player's passes that immediately precede a teammate's shot.  So there is no hope for developing a shot-based metric that is analogous to assists with data that is currently available to the public. 

However, while the NHL's historical databases do not contain information about passes that led to shots, they do contain information about the players on the ice for every shot taken, as well as the player who took the shot.  This data is exactly what is need to develop a shot-based version of altruistic contribution that is analogous to the goal-based version described in Section \ref{linemates}.  

We can define $m(A), c(A),$ and $a(A)$ in the same way we did previously, except using shots per 60 minutes instead of goals per 60 minutes.  A player's marginal contribution is found using
\begin{align*}
	m &= SF_{on} - SF_{off}.
\end{align*}
where $SF_{on}$ and $SF_{off}$ are computed like $GF_{on}$ and $GF_{off}$ using shots instead of goals.  A player's competitive contribution $c(A)$ is now defined as his shots per $60$ minutes.  The altruistic component, 
    $$a(A) = m(A) - c(A),$$ 
can be thought of as the difference in \textit{shots} per $60$ minutes by the player's teammates when he is on the ice versus off the ice.  This version of altruistic contribution using shots is what we use in the next section to develop our playmaking metric.

\section{The Playmaking Metric}  \label{improving}

We now develop our playmaking metric, which combines both assists and our shot-based altruistic contribution metric to form a measure of a player's contributions towards his teammates' productivity.  This metric has two benefits over assists which we can provide statistical evidence.  Our metric is (1) more consistent than assists, and (2) better than assists at predicting future assists.  

Points (1) and (2) are essential.  Up to this point, we have defined a shot-based version of altruistic contribution, which is a way to measure how a player affects the number of shots that his teammates take, in the same way that assists are a way to measure how a player affects the number of goals that his teammates score.  While the definition makes intuitive sense, we have not yet given any statistical evidence that these measures are actually useful or better than any existing metrics.   In this section we provide evidence that our metric is better than assists.

\subsection{Calculation and comparison}

    We compare two linear regression models: one that uses only assists as a predictor, and one that uses both assists and our shot-based altruistic contribution metric.  
    More precisely, we compare  
        \begin{equation}\label{assists-model}
                y = \beta_0 + \beta_1 A + \epsilon 
        \end{equation}
      with 
        \begin{equation}\label{play-model}
        y = \beta_0 + \beta_A A + \beta_{Alt} Alt + \epsilon, 
        \end{equation}
    where $A$ and $Alt$ denote assists and altruistic contribution per $60$ minutes in one half of a season and $y$ denotes assists per $60$ minutes in the other half of a season. 
    The expected assists per $60$ minutes obtained from \eqref{play-model} are what we call our playmaking metric.  Recall that we are only considering 5-on-5 situations in which both goalies are on the ice.

\begin{figure}[h!]
    \centering
    \includegraphics[width=.49\textwidth]{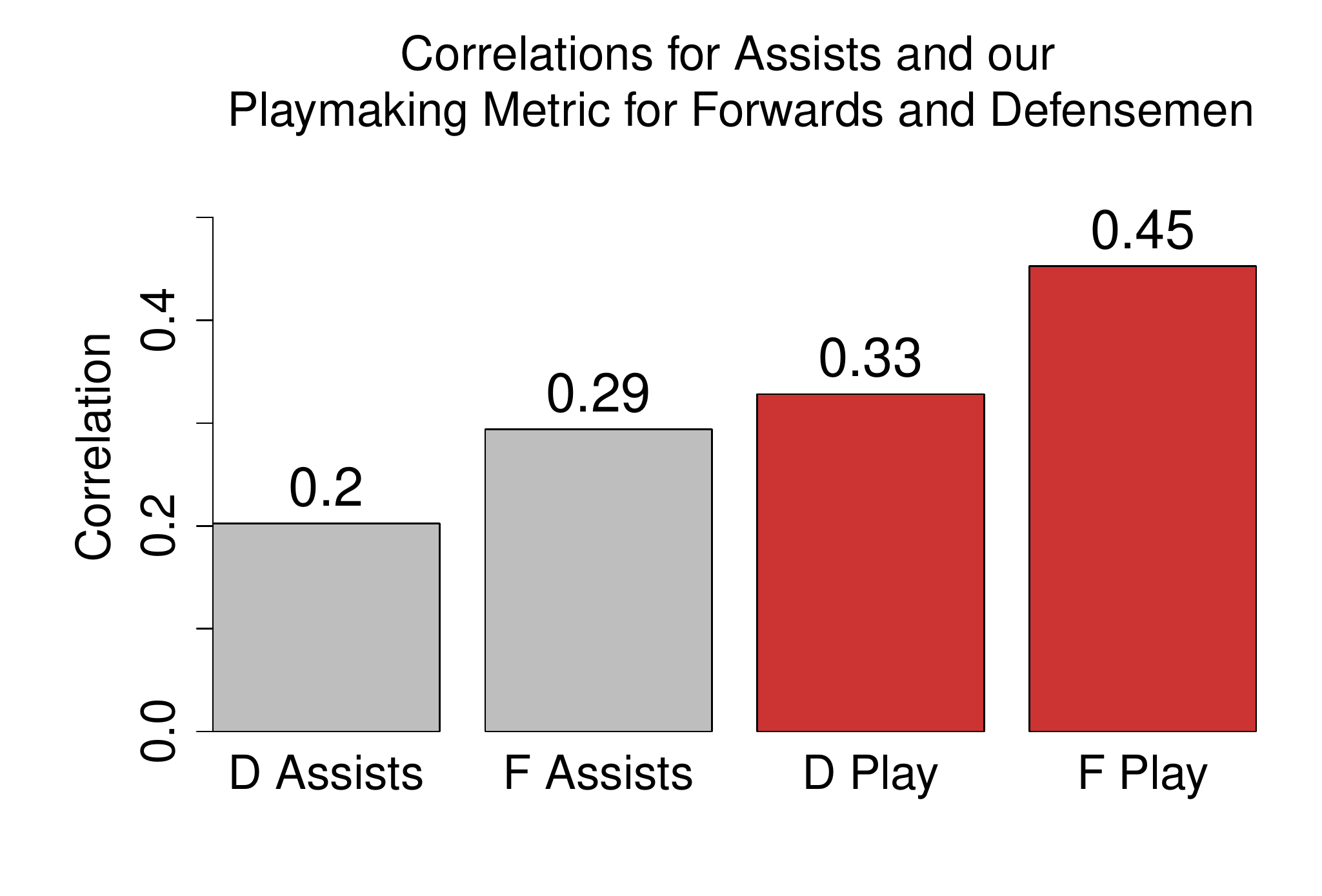}
    \includegraphics[width=.49\textwidth]{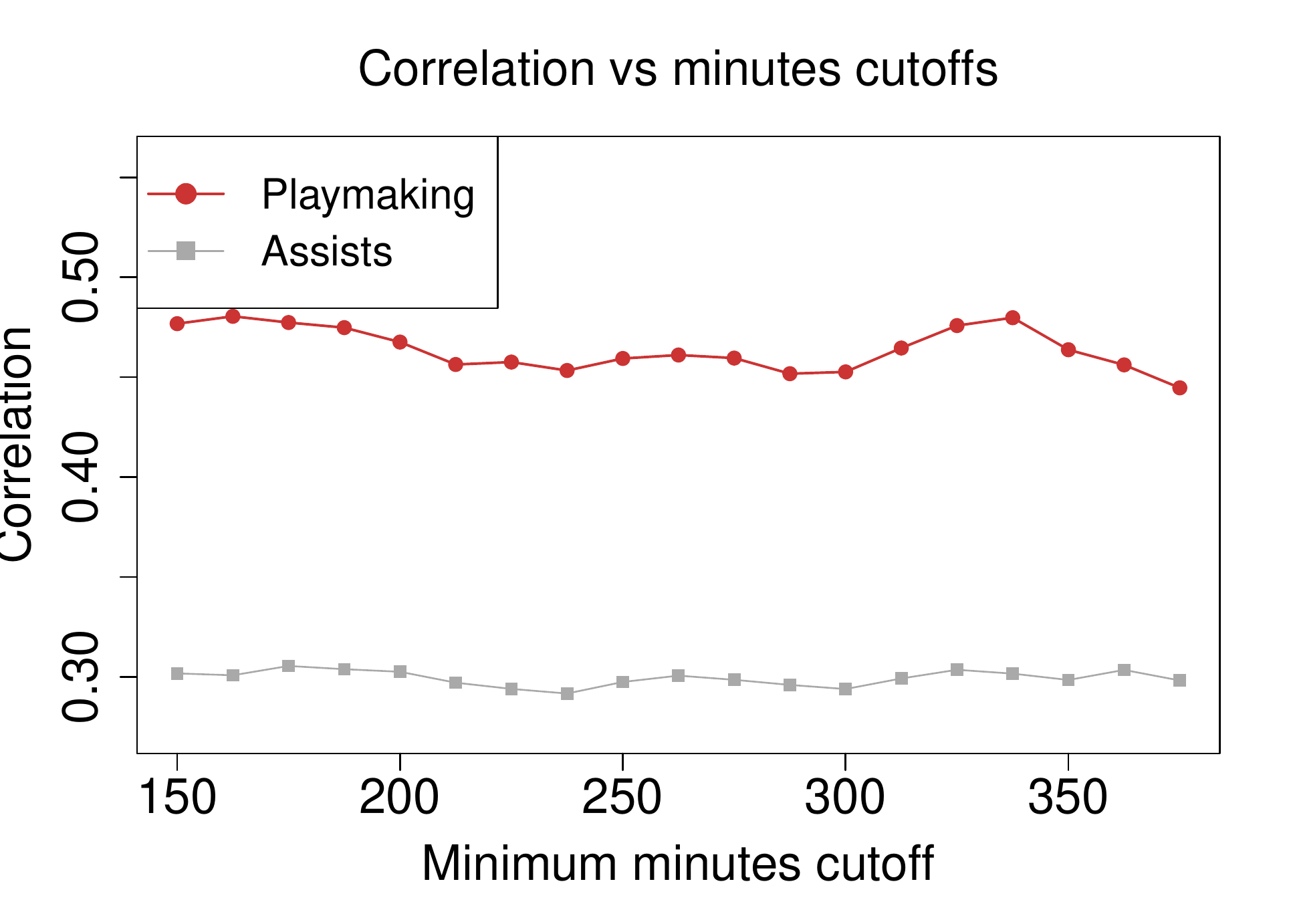}
    \caption{(Left) Half season to half season correlation of assists (gray) and our playmaking metric (red) for forwards and defensemen with a minimum of $300$ minutes of playing time in both halves of the season. (Right) Half season to half season correlations for forwards for different choices of minimum minutes cutoff.}
    \label{half-year-to-half-year-correlation}
    \end{figure}

We built these models for forwards and defensemen separately, using both half and full seasons of data.  In all cases, \eqref{play-model} outperformed \eqref{assists-model}.  Figure $\ref{half-year-to-half-year-correlation}$ illustrates playmaking is a more consistent measure of performance than assists for both forwards and defensemen.

        It is significant to note that rate statistics are vulnerable to variability for smaller sample sizes.  For example, a winger called up from the AHL to the NHL could potentially score one minute into his first NHL shift.  His resulting scoring rate would be an impressive 60 goals per 60 minutes, far exceeding that of league superstars like 2012-13 scoring leader Alex Ovechkin.  
        For this reason, we chose a minimum playing time cut-off of 300 minutes when computing these correlations.  This choice is somewhat arbitrary, so in the right of Figure \ref{half-year-to-half-year-correlation}, we show that our general conclusions do not change for different choices of cut-off.  
                                
        The results for year-to-year correlations are similar.  In particular, we get a correlation of $0.53$ for our playmaking metric for forwards. In fact, the half-season to half-season correlations for our playmaking metric are higher than the full season to full season correlations for assists. 

        Statistical measures for goodness of fit further support our playmaking metric from \eqref{play-model}.  That model has a better adjusted $R^2$, Mallows' C$_p$, and AIC than \eqref{assists-model}, which all indicate our metric is better than assists at predicting future assists.
        A $10$-fold cross-validation also showed the mean squared error of predicted assists versus actual assists is smaller for \eqref{play-model} than for \eqref{assists-model}.
        The same is true whether we divide the data into half seasons or full seasons, or use defensemen instead of forwards.

\subsection{Top playmakers}

In Table $\ref{playtable}$, we give the top five playmakers in $2010$-$11$ according to expected assists, using our full season to full season model for forwards.  These expected assists are calculated from our playmaking metric, which is in the units of expected assists per 60 minutes, along with the player's playing time that year.
    \begin{table}[h!]
    \begin{center}
    \caption{Top five forwards in playmaking ability in $2010$-$11$. }
    \label{playtable}
    {\small
    \begin{tabular}{llrrrrrrr}
      \addlinespace[.5em] & & & \multicolumn{2}{c}{$2009$-$10$} &  \multicolumn{2}{c}{$2010$-$11$} & \multicolumn{2}{c}{Difference} \\
      \toprule 
    Player & Pos & Team & A & PLAY & A & PLAY & A & PLAY \\ 
      \midrule 
      Henrik Sedin & C & VAN & $53$ & $32$ & $44$ & $31$ & $9$ & $1$ \\ 
      Anze Kopitar & C & L.A & $18$ & $22$ & $33$ & $25$ & $15$ & $3$ \\ 
      Claude Giroux & RW & PHI & $15$ & $17$ & $33$ & $25$ & $18$ & $9$ \\ 
      Daniel Sedin & LW & VAN & $36$ & $22$ & $35$ & $24$ & $1$ & $2$ \\ 
      Bobby Ryan & RW & ANA & $17$ & $19$ & $28$ & $24$ & $11$ & $5$ \\ 
       \bottomrule 
    \end{tabular}
    }
    \end{center}
    \end{table}
    The columns A and PLAY denote assists at even-strength and expected assists from our playmaking metric, respectively. The last two columns are the absolute difference between the $2009$-$10$ and $2010$-$11$ statistics.  Note that for these players, the playmaking metric tended to be more consistent from year-to-year than assists.  It is interesting that our playmaking metric had Claude Giroux as the third best playmaker in the league in $2010$-$11$, the season before he was a top three scorer.

\section{Conclusions and Future Work} \label{conclusions and future work}

In this paper, we have introduced a measure of NHL playmaking ability capable of better predicting future assists than assists themselves.  Our metric (1) adjusts for the strength of a player's linemates, (2) is more consistent than assists, and (3) is better than assists at predicting future assists.  

Identifying playmaking ability in this way can compliment the expertise of coaches, general managers, and talent evaluators to differentiate relative value among a collection of talented players.  Their decision making is assisted by better identifying and understanding the potential value of a prospective player joining their organization.  Trade targets, free agent signings, and draft picks can be assessed by not only using traditional performance measures, but also through considering their fit into the chemistry of an existing organization.  Specialization in terms of playmaking ability, competitive contributions, and altruistic contributions can be targeted in accordance with a team's needs.

Several possibilities exist for future study.  Alternative measures of player marginal contributions can be explored using the player ratings 
in 
\cite{thomas-ventura}, 
\cite{schuckerscurro}, and 
\cite{gramacy-jensen-taddy},
or the adjusted plus-minus ratings 
in \cite{apm}, \cite{apm2}, and \cite{ridge}.  These choices of marginal contribution may be preferred since they account for the strength of a player's opponents, and in some cases, the zone in which a player's shifts typically begin.  Additionally, although we focus on contributions within a competitive sports team, similar analysis could benefit any team, organization, corporation, or military unit working towards a common goal, albeit quantifying such scenarios is difficult in the absence of well defined value and payoff functions available in competitive sports.  

Lastly, we note that our focus was on the case where $A$ denotes a single player, since we were most interested in developing a metric for an individual player's playmaking ability.  However, all of the definitions of marginal, competitive, and altruistic contributions remain the same in the case where $A$ is a subset of two or more players.  In this case, an assessment of chemistry between two or more teammates can be pursued in an attempt to reveal what player combinations yield higher on-ice productivity.
 
\bibliographystyle{DeGruyter}
\bibliography{playbib}

\end{document}